# Towards Unsupervised Learning based Denoising of Cyber Physical System Data to Mitigate Security Concerns

Mst Shapna Akter*, Hossain Shahriar†

*Department of Computer Science, Kennesaw State University, USA
† Department Information Technology, Kennsaw State University, USA
{* makter2@students.kennesaw.edu,† hshahria@kennesaw.edu }

*Abstract*—A dataset, collected under an industrial setting, often contains a significant portion of noises. In many cases, using trivial filters is not enough to retrieve useful information i.e., accurate value without the noise. One such data is time-series sensor readings collected from moving vehicles containing fuel information. Due to the noisy dynamics and mobile environment, the sensor readings can be very noisy. Denoising such a dataset is a prerequisite for any useful application and security issues. Security is a primitive concern in present vehicular schemes. The server side for retrieving the fuel information can be easily hacked. Providing the accurate and noise free fuel information via vehicular networks become crutial. Therefore, it has led us to develop a system that can remove noise and keep the original value. The system is also helpful for vehicle industry, fuel station, and power-plant station that require fuel. In this work, we have only considered the value of fuel level, and we have come up with a unique solution to filter out the noise of high magnitudes using several algorithms such as interpolation, extrapolation, spectral clustering, agglomerative clustering, wavelet analysis, and median filtering. We have also employed peak detection and peak validation algorithms to detect fuel refill and consumption in charge-discharge cycles. We have used the R-squared metric to evaluate our model, and it is 98 percent In most cases, the difference between detected value and real value remains within the range of ±1L.

## I. INTRODUCTION

Nowadays, new features are introduced to the connected vehicular environment. The latest feature is connecting the vehicle's sensor with a smartphone application via bluetooth or WiFi etc. Some applications create easy access for attackers to collect information. They can misuse the system for their own purposes, which can be harmful to vehicles and passengers. Cyber-attacks can be generated via wireless systems which transmit a large volume of information from the vehicle, for instance, internet, wifi, or bluetooth. However, cyber security experts have limited exposure to certain security systems such as Contoller Area Network (CAN). The involvement of CAN bus attacking codes in a mobile malware is cost-effective for the malicious attackers but very challenging for vehicle engineers to alleviate the security risks [1]. Moreover, many hazards are identified based on airflow, fuel, RPM and ignition. CAN protocol is used for scheduling analysis to find out whether the messages are transmitted in a timely manner or not. In this scenario, noisy data may cause complexity in the system. Noisy data can be injected to perform attacks on cyber physical system, so denoising is an important step. In this work, the focus revolves around the industrial dataset that has an issue of a severely noisy environment and security issues. All the vehicles have been operative on the street in Bangladesh and the data collection involves wireless technology, including Bluetooth and internet server. Due to the poor condition of a few roads in Bangladesh, noise may have been prominent in the dataset, which can cause the worst situation in the cyber security domain as a noisy dataset contains improper information and can be modified further for attacker's own purpose. Therefore, it is necessary to denoise the information to present its actual form. Data denoising or removing noise from a dataset is a challenging topic for the researchers. Each dataset is different in terms of data distribution and presence of noises. These noises are generated due to device malfunctioning, human limitation, machine limitation, improper approaches of collecting and preserving data, etc. Barely any work has been done in denoising time-series data containing fuel information retrieved from the industrial domain. Since this is time-series data, any given point within a certain time frame either indicates charging or discharging. Noise present in the dataset gives a false representation of peak or consumption. Here, false representation stands for noisy data that plagues the actual value. when the engine or machine is switched off, no charging or discharging occurs within a certain period. This yields a stationary or constant set of values within that time frame. Our system has required to deal with this issue, and it has accomplished successfully for most of the parts. Since the dataset does not contain the labels, we have used the unsupervised approaches. Our system has started with simple interpolation, extrapolation, and white noise removal techniques. Here, by white noise, we are referring to random noise present in a dataset whose mean value is zero and has a finite variance. Then two clustering methods have been brought to isolate noise from real value. Wavelet analysis has cleaned it further through wavelet transformation and inverse transformation. Finally, median filtering has been used to give the final push. At first, we have trained our model using a concatenated dataset containing information from multiple vehicles. Then, to test it further we have been provided with an additional dataset of various vehicles. The initial dataset contains over 100,000 data points, which have been narrowed down to 37 peak values. To evaluate our result properly, we have used the R-squared metric and compared our result with previous work that has been done on similar (but certainly not identical) datasets [2]. The R-squared value of the final result is 0.98, and RMSE is 1.49. Our system detects not only the refill of fuel but also the consumption rate. The rest of the paper is organized as follows: In Section II, we introduce the related work of data denoising approaches. Section III describes the methodology of our proposed techniques. The experimental results are explained in Section IV. Finally, Section V concludes the paper.

## II. RELATED WORK

One area where denoising techniques have been implemented is audio signals [3]. Audio signals usually come from speech, background noises, or any other automated source (electronic, electromagnetic, acoustic). The suggested methodology is heavily dependent on filters such as Kalman filter, Butter Worth, Chebyshev, Elliptical Filters, and so on. In the medical sector, ECG (electrocardiogram) signals are also cleaned using filters such as low pass filter [4]. Apart from these, deep learning methodologies [5] and non-linear diffusion [6] have also been considered. We have already ruled out a deep learning technique due to the unsupervised nature of the data [7, 8], which we have dealt in this work. The other problem is that: relying just on filters produces poor results due to high regularization. In contrast, our model tends to be generic, which means it will work efficiently on any vehicle with ambiguous noise patterns. It is evident that: the contemporary techniques are focused on denoising images or audio signals. Although denoising techniques are applied in the realm of IoT (internet of things) by using ANN (artificial neural network) [9, 10], ToF (Time-of-Flight) data by using GAN [11], and vibration sensor data by using TFM (time-frequency manifold), none of the solutions seems suited for our dataset due to unsupervised nature of the data. In our case, we had to tackle two parallel problems; one is to minimize data loss, and another is to get the actual value in a completely unsupervised manner.

## III. METHODOLOGY

All the data have been collected from Pi Labs Bangladesh Ltd. Initially, from 9 vehicles, roughly 1,00,000 data points have been collected and used for our system. Data flow is shown in Fig.1, which demonstrates the cycle starting from raw data to processed data. It begins with employing industrial settings in a moving vehicle. Then some generic transmission occurs. Our algorithm can reduce noise from the time-series data while it is streamed from source to destination through a server.

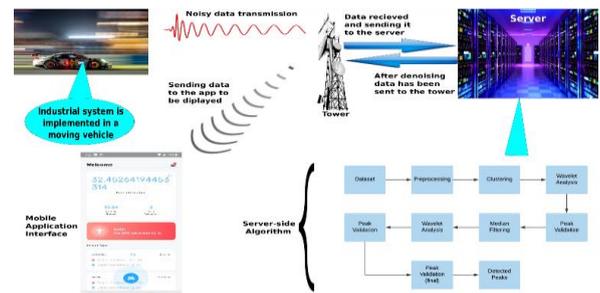

Fig. 1: A general overview of the data cycle. The industrial setting is implanted in a vehicle, and it collects and sends data to the server. Inside the server, our algorithms are implemented where the data traverse through each phase. After successful peak detection, the data is sent to the mobile application (beta version) from where the result can be displayed.

We wanted to test whether real-time data can be displayed, and this is why a mobile application prototype has been built. Our system begins cleaning the dataset with simple extrapolation and interpolation. This removes some trivial types of noises i.e., white noise. Then two sets of clustering methodologies (spectral and agglomerative) isolate noise from the actual value. After that, the dataset is stored and sent to the next phase to perform wavelet analysis. This generates two datasets to be compared and keep the common values. A similar strategy is undertaken for median filtering and wavelet analysis. Finally, when all prior operation is done, we have performed a final validation to get the peak values and consumption rate. Fig.2 (a) gives a generic outlook of the data set.

### A. Data Preprocessing

Initially, some rows contained zero values. This could contaminate the results of further processes, and therefore data needed to be tackled in an efficient manner where data replacement conforms to the pattern of non-zero values.

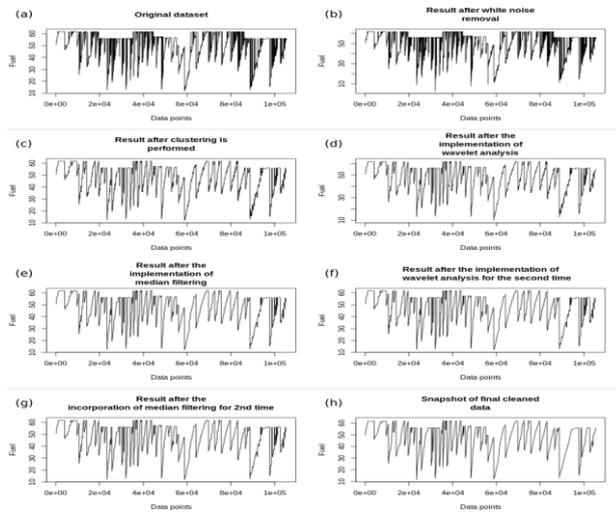

Fig. 2: A general overview of the data cycle. The industrial setting is implanted in a vehicle, and it collects and sends data to the server. Inside the server, our algorithms are implemented where the data traverse through each phase. After successful peak detection, the data is sent to the mobile application (beta version) from where the result can be displayed.

Preprocessing involves white noise removal using various data wrangling techniques and applying interpolation-extrapolation to recover the lost points. Also, we have differentiated the dataset to create a new feature. Interpolation is a method that can be achieved using discrete data set and by forming general formula within a certain range. In our case, linear interpolation was used, which can be expressed as $y = y_1 + (\frac{y_2-y_1}{x_2-x_1})(x-x_1)$ Here, unknown points are denoted as y for a certain value of x. Extrapolation operates outside of the range of the observation. It is subjected to higher uncertainty, as opposed to interpolation, since it fills out data based on the relation of other data points. In this research, midpoint extrapolation was incorporated, denoted as $(x, y) = (\frac{x_1+x_2}{2}, \frac{y_1+y_2}{2})$ which is operative for one unknown point. Interpolating and extrapolating data points were achieved through the following process. At the beginning, it is checked whether the data points contain 0 values or not. If it does so, all those values are set to NULL. Finally, through interpolation and extrapolation, new data has been generated and is being stored. Apart from that, a general form of noise also known as random noise, has been removed from the data set. The main characteristic of the noise includes the equal intensity of the signal at different frequencies. If the mean equals zero, then data is considered white noise, as data is spread across the negative and positive portion of the graph in an equal manner. The white noise is excluded from the dataset by using the following formula iteratively: $\alpha = (x_i - x_{i+1})(x_i - x_{i-1})$. However, there is a strict condition for white noise removal, which is $x_i \neq x_{i+1}$, and $x_i \neq x_{i-1}$. The iterative process can be represented as $y_i = x_i\alpha$. After successful completion of preprocessing, we have gotten a visual representation of the data set, as presented in Fig.2 (b).

*B. Clustering*

Clustering is an unsupervised learning method. It is used to group data according to their respective characteristics. Any data in a single cluster possesses a similar characteristic as every other data point in the same cluster. Right after the preprocessing is done, two clustering techniques have been incorporated. These clustering methods work together to separate noisy data from correct data. Spectral clustering is best suited when the variation of data is not much. It clusters the data based on the density of data points. The state of being close proximity to each other is called affinity, and this phenomenon can be described by the affinity matrix. Different vectors from this matrix can be extracted using principal component analysis (PCA), which later leads Eigenvectors to be formed. These vectors are referred to as feature vectors of each object of affinity or Laplacian matrix. Hierarchical clustering, also known as Agglomerative clustering, generates a tree where roots are the lowest point of the data-set, and leaf nodes consist of values that are greater than the values of the root node. It uses the bottom-up approach to group the data points in a hierarchical manner. Every data point groups together in its own cluster. This goes on for all data points, and these clusters are then joined using a greedy approach. The greedy approach involves merging two most similar clusters together. To perform the clustering on the entire dataset we have incorporated the following formula $Y_{s,h} = \alpha f_s(x_i) + (1-\alpha)f_h(x_i)$ here, $f_s$ and $f_h$ represent the mechanism of spectral and agglomerative clustering respectively as functions. At the start of clustering, we set a constant value denoted as threshold $T$ whose value is 0.1. Then standard deviation is being calculated based on all the data points. As agglomerative clustering uses a dendrogram to cluster that data it works well on the dataset having higher deviational value. As opposed to that, spectral clustering checks on the affinity of the data points to group them; That is why it is better suited for dataset having lower deviational value. Due to variation of standard deviation after each iteration, we used two clustering. If $T$ is greater than standard deviation, then $\alpha = 0$ and data points are grouped using Agglomerative clustering, and if it is less then, $\alpha$ is set to 1, and data are grouped together using spectral clustering. Each iterative step is incremented by 200. After clustering, many data points get eliminated as part of the noise removal technique. Interpolation and extrapolation are used to fill out those data points. Fig.2

(c) is a visual representation of the effect of clustering on the data set.

## C. Wavelet Analysis

Unlike any other Fourier transformation methodology, wavelet transformation is used for transforming data represented in the time domain to the frequency domain. In previous work, we have found that it has the ability to compress an image efficiently. By managing factors like shifting and scaling, it can decompose an image to multiple lower resolution images. The waves have features like varying frequency, limited duration, and zero average value. This is also eligible to remove high-frequency noise from any time-series data with non-continuous peaks. The implementation of wavelets revolves around implementing two different transformations and incorporating one threshold function. These transformations are wavelet transformation and inverse wavelet transformation. The wavelet transformation is achieved by the following formula:- $C_{\tau,s} = \frac{1}{\sqrt{s}} \int_t f(t) \Psi(\frac{t-\tau}{s}) dt$ Above is the formula for continuous wavelet transformation where $\tau$ and $s$ are transition parameter and scale parameter respectively, $\frac{1}{\sqrt{s}}$ is normalization constant, and $\int_t \overline{f(t)} \Psi(\frac{t-\tau}{s}) dt$ is the mother wavelet. The inverse operation is carried out by the following function:- $f(t) = \frac{1}{\sqrt{s}} \int_\tau \int_s C(\tau, s) \psi(\frac{t-\tau}{s}) d\tau ds$ Discrete wavelet transformation is a bit straight-forward than this and, to state the obvious, free from the integral operation. The formula for discrete wavelet transformation is $a_{jk} = \sum_t f(t) \psi_{jk}(t)$ and inverse discrete wavelet transformation can be written as such $f(t) = \sum_k \sum_j a_{jk} \psi_{jk}(t)$, . These formulas provide simultaneous localization in time and scale, sparsity, adaptability and linear time complexity; which allow noise filtering, image compression, image fusion, recognition, image matching and retrieval efficiently. Finally, an additional threshold function can be represented as such to improve the proposed model, $threshold = alpha * \sqrt{noise} * log(datasize)$ and in this formula, alpha is a constant and noise = absolute median value. We incorporated the discrete approach in our model. This reserves the original signal but without the noise. After the implementation, the peak value was shifted by 5200 points. Fig-2 (d) and (f) show the result after wavelet analysis.

## D. Mediun Filtering

Median filtering is a non-linear filtering technique, and it removes noise from the data set while preserving the valuable data. In our case, the median filter is to traverse through the signal one by one. It generates a window to slide through the data entry, and this window is generated based on the pattern of the neighboring window. This replaces each entry with the median of neighboring entries and what remains is the peak. Fig.2 (e) and (f) show results after median filtering applied for the first time on the data set. Our data has one dimension. A window of the median filter contains the first few preceding and following entries. To carry out this task, we used this formula, $Y(X_{top}, X_{bottom}) = M(x_i, x_{x+1})$, where $M = median$, $X_{top}$ = Initial point, $X_{top}$ = Last point, and $x_i, x_{i+1} \epsilon X$. Once what algorithm to use is sorted out, we have been required to write our own custom algorithms where we can use clustering, filtering methodologies efficiently. It is necessary to choose data window and iterative steps carefully so that we can minimize the data loss. Our custom algorithms include one peak detection and three peak validation methods.

## E. Peak Detection

This is the part where peaks are detected. The algorithm uses the moving average to find any sudden changes in the data points. As we pass through data points, we calculate the simple moving average with $SMA = \frac{\sum x_i}{3}$. We used a window of 3 points to find any discrepancy in the data. So, for every three adjacent point $SMA = \frac{\sum_{i=1}^{3} x_i}{3}$. Any deviation of more than 4 units is considered a peak. There are three peak validation methods in this system, among which two are almost identical. These validation methods validate the data points, whether they are valid peak points or not. The prior two peak validation methods include the following two sections.

## F. Peak validation (first and second)

At the start, peak validation data is sent to two different directions. One part is operative using clustering and peak detection algorithm. Another part is operative using a combination of clustering, wavelet analysis, and peak detection algorithm. This produces two different datasets. After that, each entry of one dataset is evaluated against the corresponding entry of other data set. If there is a similarity, we validate the peaks; otherwise, we discard it. This parallel processing gets the work done fast without interrupting each other. The formula for peak validation (first) is $f_{rp} = |f_{pcp} - f_{pwp}| \leq 100$, where $f_{rp}$ = valid peak points, $f_{pcp}$ = = data points obtained after performing peak detection on clustered data points = $SMA(Y_{s,h}(X_i))$ and $f_{pcp}$ = data points obtained after performing peak detection on data points retrieved from implementation of clustering and wavelet analysis $W_A = SMA(W_A(Y_{s,h}(X_i)))$. In order to get a valid peak $f_{rp}$ must be less than or equal to 100. It is a comparison between values obtained by only after clustering and a combination of clustering and wavelet. Real peaks are validated based on similarity. The second peak validation method, as mentioned earlier, works similarly. The only difference is that, in second peak validation, median filtering has been used instead of clustering. Likewise, it is a comparison between the two datasets. One is obtained after performing just filtering, and another is acquired through implementing a combined system of filtering and wavelet. So, the formula for second peak validation is, $f_{rp} = | f_{pcp} - f_{pmp} | \vee \leq 100$ where $f_{pcp}$ = data points obtained after performing peak detection on data points retrieved from the implementation of clustering and median filtering $SMA(M_F(Y_{s,h}(X_i)))$. As part of the peak validation, the effect of wavelet transformation and median filtering, which is implemented for the second time, can be found in Fig.2 (f) and Fig.2 (g), respectively.

*G. Peak validation (final)*

This peak validation method is operative on the final data set before producing the final result. Three consecutive data points, previous = i-1, current = i and next = i+1, are considered for data traversal. We have defined two variables, such as A = previous – current and B = current – next, to check the distance and difference of A and B. Distance refers to the position between A and B, and the difference is the value obtained by subtracting A and B. If the distance is less than 30 and the difference is less than 5, then the value is removed from the peak. This final procedure isolates the peak from the rest of the data to have a proper evaluation of peak detection. After this peak validation, we obtain 37 peak points. After getting the peaks, we can easily calculate the consumption rate by finding the difference between peaks. This final procedure isolates the peak from the rest of the data to have a proper evaluation of peak detection.

## IV. RESULT AND DISCUSSION

Our paper has focused primarily on noise reduction and peak detection from the industrial dataset retrieved from vehicles. To achieve this target, we have faced an intricate trade-off; detecting peaks while minimizing data loss. We have implemented different algorithms mentioned in the previous section to get the actual value from the noisy time series dataset. We have achieved very accurate results by mashing up different algorithms according to the type of noise. Our two-layered noise reduction method has detected peaks while eliminating the noise based on a threshold value.

Table-1: List of detected peaks as compared to the real value. For the most part, the error is minimum.

As seen in Table-1, for the most part, our module can detect peaks with an error of ±1L. Those values whose error rate is higher than 5 liters are due to the reason mentioned above. However, the R-squared score is 0.98, and RMSE is 1.49, which indicates that our module is highly efficient. The graphical representation of cleaned data after all noise removal can be seen in Fig.2 (h).

Due to hardware malfunction or some similar reason, at different time periods, the obtained data showed bizarre patterns, which has caused discrepancy in finding the correct result. In the highway, due to the high speed, the fuel consumption rate remains constant, as the device has its limitation. As a result, there is a huge data loss during this time period. If the dataset collected has no data loss, then the result can be significantly improved where the accuracy may exceed ours.

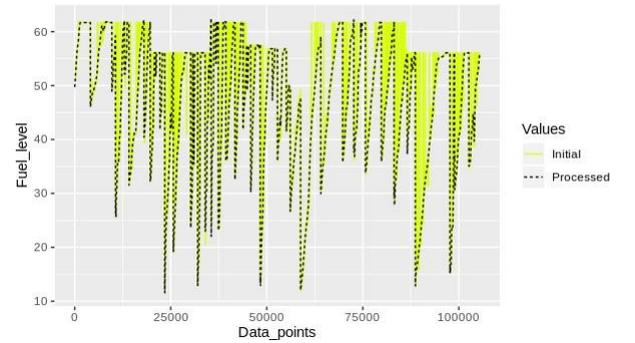

Fig. 3: A comparison between our initial dataset and processed dataset. The data points are narroweddown, eliminating all the noises that were present at the beginning.

ACKNOWLEDGEMENT

The work is partially supported by the U.S. National Science Foundation Award #2100115. Any opinions, findings, and conclusions or recommendations expressed in this material are those of the authors and do not necessarily reflect the views of the National Science Foundation.

## V. CONCLUSION

In conclusion, the real dataset (on which we have worked) has contained severe noise, which may have occurred due to poor street condition of some roads of Bangladesh, noisy dynamics, mobile environment and so on. The proposed module in this work, regardless of the intensity of data noise, is capable of detecting peaks without removing the noise in the first place. The reason behind the capability of doing that: different methods have been arranged in such a way that it can store the data before sending it to the next phase or process. For instance, data generated from clustering is stored first before sending it to the following phase to perform wavelet transform. This strategy allows the system to compare the values of the datasets retrieved from two adjacent processes. Before removing the noise, this entire task is conducted using multiple peak validity and peak detection methods. Consequently, the system works more efficiently as it is independent of the noise percentage. So regardless of the percentage of noise present in the datasets, peaks have been detected. However, after successful peak detection, the noise has been gotten removed and the correct value of the consumption rate has also been generated. So, this system can isolate noise and peak first; and then remove noises from the rest of the datasets, which provides not only correct peak values but also consumption values as well. Since this is a two layer-based noise removal system, at first, peaks have been separated from the noisy dataset. Then, the consumption rate is deduced using these separated peaks. This approach helps to maintain the integrity of the peaks. This type of design allows us to minimize data loss during noise removal, which has not been reported previously in the presence of bizarre data patterns i.e. constant values, accuracy subsides. Without such extreme cases, accuracy may increase a lot. Our work can be implemented in any of the industrial fields, where exist the complicated issues of fuel measurement i.e. vehicle industry, fuel station, power plant and so on. Moreover, removing the noisy part may ensure less cyber security attacks in vehicular systems. As long as any time-series data distribution resembles our dataset distribution, the module can work efficiently to reduce noise and produce correct values.